\documentclass[apj,twocolumn,numberedappendix]{openjournal}
\usepackage{newtxtext,newtxmath,enumitem,multirow,ctable}
\usepackage[T1]{fontenc}
\usepackage[breaklinks,citecolor=blue]{hyperref}
\hypersetup{
    colorlinks=true,
    linkcolor=blue,
    filecolor=blue,      
    urlcolor=blue,
}
\usepackage{lipsum}
\usepackage{rotating}

\newcommand{\etal}{et al.}

\shorttitle{TRAPPIST-1 external planets' emission}
\shortauthors{Cartigny \etal}

\begin{document}
\title{Looking for TRAPPIST-1 external planets' emission in JWST archival data}
\author{Louis-Julien Cartigny}
\affiliation{Institut d’Astrophysique de Paris, Sorbonne Universit\'e, CNRS, 98 bis bd Arago, 75014 Paris, France}
\affiliation{LIRA, Observatoire de Paris, Universit{\'e} PSL, CNRS, Sorbonne Universit{\'e}, Univ. Paris Diderot, Sorbonne Paris Cit{\'e}, 5 place Jules Janssen, 92195 Meudon, France}

\author{Alice Maurel}
\affiliation{Institut d’Astrophysique de Paris, Sorbonne Universit\'e, CNRS, 98 bis bd Arago, 75014 Paris, France}
\affiliation{Laboratoire de M\'et\'eorologie Dynamique/IPSL, CNRS, Sorbonne Universit\'e, Ecole Normale Sup\'erieure, Universit\'e PSL, Ecole Polytechnique, Institut Polytechnique de Paris, 75005 Paris, France}
\author{Elsa Ducrot}
\affiliation{AIM, CEA, CNRS, Universit\'e Paris-Saclay, Universit\'e de Paris, F-91191 Gif-sur-Yvette, France}
\affiliation{LIRA, Observatoire de Paris, Universit{\'e} PSL, CNRS, Sorbonne Universit{\'e}, Univ. Paris Diderot, Sorbonne Paris Cit{\'e}, 5 place Jules Janssen, 92195 Meudon, France}

\author{Taylor J. Bell}
\affiliation{AURA for the European Space Agency (ESA), Space Telescope Science Institute, 3700 San Martin Drive, Baltimore, MD 21218, USA}

\begin{abstract}
The TRAPPIST-1 system has been thoroughly observed with JWST. Unfortunately, stellar contamination issues strongly limit the interpretation of transit observations. As for emission observations, only the two closest planets have been observed through five dedicated JWST programs. We gathered all these emission observations and tried to detect the combined emission of the external planets in previous JWST MIRI observations of TRAPPIST-1; this paper presents our approach and our results. We could not achieve sufficient precision to detect the thermal emission of the outer planets and to discriminate between an all-bare rocks and an all-atmospheres scenario. However, we show that a $\simeq$60~hours continuous observations at a specific phase range, when the gradient of thermal flux  from the outer planets is maximum, could allow us to achieve this goal.
\end{abstract}
\maketitle

\section{Introduction}\label{sec:intro}

Since the discovery of the first exoplanet around a main-sequence star \citep{mayor:1995},  more than 6,000~exoplanets have been detected, with several thousands more yet to be confirmed (\hyperlink{https://exoplanetarchive.ipac.caltech.edu/}{NASA Exoplanet Archive}). After two decades mainly focused on detecting and studying giant planets with short orbital periods---due to the observational biases imposed by the main detection methods being the transit and radial velocity methods---it is now possible to study rocky, temperate planets. The rocky planets that can currently be characterized orbit mostly around the coolest and smallest stars, M-dwarfs \citep{nutzman_charbonneau:2008, barclay:2018, triaud:2021}. The question of the existence of atmospheres on such exoplanets is one of the key subjects of interest in exoplanetology. While M-dwarfs are good targets for exoplanet detection and characterization, they also have a high early-life activity expected to drive substantial atmospheric escape \citep{bolmont2017, bourrier:2017}. The question remains if secondary atmospheres can be formed and maintained on such planets \citep{krissansen-totton:2024}. To date, the competition between the planet's capability to retain its atmosphere and the star's tendency to trigger atmospheric escape is embedded in the concept of a ``cosmic shoreline" \citep{zahnle:2017}.

The James Webb Space Telescope (JWST) possesses unique capabilities in the infrared, necessary to put constraints on the atmospheres of rocky exoplanets. Unfortunately, the most common method of detecting atmospheres, transit spectroscopy, is strongly contaminated by the Transit Light Source (TLS) effect for M-hosts, strongly limiting the exploitation of transit observations \citep{rackham:2018,lim:2023}. Fortunately, JWST can also observe the mid-infrared emission of rocky exoplanets that are warm enough, through secondary eclipse photometry and spectroscopy (for the planets with the highest Emission Spectroscopy Metric, ESM \citep{kempton:2018}). This technique has been used on several planets, and it is the one chosen for the Rocky Worlds DDT, which includes 500-hours of JWST observations aimed at looking for atmospheres on rocky exoplanets and probing the cosmic shoreline \citep{redfield:2024}.

The two innermost planets of the TRAPPIST-1 system (b \& c) \citep{gillon:2016} were among the first rocky planets to be observed in emission with JWST (GTO~1177,1279). This system, at only 40~light years from the Solar System, is composed of at least seven Earth-sized transiting planets orbiting an M8-type star with short orbital periods ranging from 1.5 to 18 days. Three of these planets are in the habitable zone, which makes this system a very interesting target not only for exoplanetology but also for the understanding of habitability. This system has been thoroughly studied and observed and was the subject of eleven JWST programs so far, including five in emission. The eclipse observations of TRAPPIST-1~b \citep{greene:2023, ducrot:2024} suggested that it does not hold a substantial atmosphere, and similar observations for c \citep{zieba:2023} still leave room for atmospheric scenarios. A phase curve of the system has also been observed \citep{ducrotgillon:2024}, confirming the conclusion that TRAPPIST-1~b is most probably airless, and putting additional constraints on potential atmospheres around TRAPPIST-1~c. Due to their longer orbital periods, it is not possible to perform the same kind of study in emission for the five outer planets, as they are cooler and would need a very large amount of observation time from JWST. However, the outer planets are also more likely to have an atmosphere \citep{krissansen-totton:2024}.

Although the outer planets are too cold to be detected individually in emission, the question remains if their combined emission could be observed. In this work, we re-analysed all JWST emission observations of the TRAPPIST-1~system to see if the combined emission of the outer planets can be characterized. We modelled the total emission of the system for two opposite scenarios---all airless planets, and full heat redistribution for the five outer planets---in order to try and distinguish between the two possibilities.

\section{Methods}
Our general strategy was to model a global phase curve of the system with the contribution of every planet, and then to compare every observation to this modelled phase curve to assess if there are some observations where the contributions of the planets sum up and could be detectable.

\subsection{Phase curve models}\label{sec:phase_curve_models}
For this project, we use very simple ``atmospheric" models, since we do not need a fine-tuned modelling of the situation but rather an estimation of the timing and amplitude of the contributions of the planets.

\paragraph{Thick atmospheres}
For the scenario where all outer planets have thick atmospheres, we consider that each planet produces a flat phase curve, with a flux corresponding to a blackbody at the equilibrium temperature for a zero-albedo planet: $$T_\mathrm{eq}=(1/4)^{1/4}\sqrt{\dfrac{R_*}{a}}T_*,$$
where $R_*$ is the star's radius, $a$ the planet's orbital semi-major axis, and $T_*$ is the star's effective temperature.

\paragraph{Airless planets} 
For the airless case, we suppose a case with no heat redistribution and a zero albedo. In this case, 
$$T_\mathrm{day}=(2/3)^{1/4}\sqrt{\dfrac{R_*}{a}}T_*,$$ \citep[see][for the demonstration]{koll:2022, maurel:2025-manuscript}. This gives the maximum of the phase curve of each planet, while the minimum of the phase curve is zero, and the phase curve variations are modelled as a simple sinusoid.

\paragraph{Relative flux}
We then compute the planet-to-star flux ratio:
\begin{equation*}
    \frac{F_{p}}{F_{*}}=\left( \frac{R_{\rm p}}{R_{*}} \right)^2 \frac{\int_{\lambda_1}^{\lambda_2}\mathcal{B}_{\lambda}(T_{\rm p})d\lambda}{\int_{\lambda_1}^{\lambda_2}\mathcal{B}_{\lambda}(T_{*})d\lambda},
\end{equation*}
where $F_{\rm p}$ and $F_*$ are respectively the planet's and star's fluxes, $R_{\rm p}$ and $R_*$ the radii, $T_{\rm p}$ and $T_*$ the temperatures, and $B_\nu(T)$ corresponds to the the flux of a blackbody at a given temperature and wavelength.

\paragraph{Eclipses and transits}
The eclipses are modelled by setting the emitted planetary flux to zero while the planet would be behind the star from our perspective. The transits are not modelled, as we are focusing on emitted flux.

\paragraph{TTV implementation}
Gravitational interactions between planets induce transit (and eclipse) timing variations (TTVs and ETVs). These TTVs are well modelled \citep{agol2021} and are taken into account for each observation by correcting the timing of the transit closest to the observation with the corresponding predicted TTV from the model from \cite{agol2021}.

\subsubsection{The \emph{\texttt{Exoplanets\_Phase\_Curves}} code}
The code for this modelling is available on Github: \hyperlink{https://github.com/LJ-Cartigny/Exoplanets_Phase_Curves}{\texttt{Exoplanets\_Phase\_Curves}} \footnote{\url{https://github.com/LJ-Cartigny/Exoplanets_Phase_Curves}}. It can model the phase curve of a specified TRAPPIST-1 planet or the total flux of a set of planets among the seven, either in relative flux or in absolute flux, at a given time for the F1280W and F1500W MIRI filters. It can consider both the airless and thick atmospheres cases discussed in \S\ref{sec:phase_curve_models}, using the formalism of \textcite{cowanModelThermalPhase2010}. As we will discuss in \S\ref{sec:obs_predictions}, the code can also be used to predict when the variation of the total flux of a given set of planets will exceed a certain threshold, which can be useful to plan observations.

The phase of each planet is modelled as a sinusoid: $$g(t)=\frac{1}{2}\sin{\left(\frac{t-t_{transit}}{P}2\pi\right)}+\frac{1}{2},$$ where $P$ is the planet's orbital period and $t_{transit}$ is the time of transit. The code looks for the nearest transit of each planet from the given initial time in the predictions of \textcite{agol2021}, which is then used as a starting point to compute the sinusoid over a given duration. The code can only model eclipses for now (because, as discussed in \S\ref{sec:phase_curve_models}, our study focuses on eclipses), taking into account the TTV predictions from \cite{agol2021}, but it could technically model transits too.
To compute the stellar flux, the code can use the SPHINX stellar spectrum library \citep{iyer:2023} for both the relative and absolute flux, or the PHOENIX library \citep{allardModelsVerylowmassStars2012} for the absolute flux.

The code focuses on the TRAPPIST-1 planets for now but could easily be adapted to other planetary systems, as it only requires several parameters: the mass, radius, and a spectrum model of the star; the planets' orbital semi-major axes, orbital periods, eccentricities, inclinations, arguments of periastron, and radii; the bandpass of the filter used; and, if we want the absolute flux in mJy, the distance between the planetary system and the JWST. Documentation is available on \hyperlink{https://lj-cartigny.github.io/Exoplanets_Phase_Curves/}{GitHub Pages} \footnote{\url{https://lj-cartigny.github.io/Exoplanets_Phase_Curves/}}.

\subsection{Data reduction}
To compare the data to our phase curve model, we need to be able to compare the flux of the system (star + planets) at the different visits. Thus, contrary to the usual data reduction in secondary eclipse or phase curve observations, we could not directly use a relative flux $F_p/F_*$ computed independently at each visit, since we precisely need to assess the difference in the flux of the star plus the outer planets (commonly called ``stellar flux" in other works) across different visits.\\
For this, we used two alternative methods. The first one consists of comparing the absolute ``stellar fluxes" (\S\ref{sec:meth/calib_stellar_flux}), i.e. the measured absolute flux during the eclipses at the different visits. The second consists in performing a global analysis of all the observations together, treating the data like one single observation (\S\ref{sec:meth/global_reduc}).

\subsubsection{Calibrated stellar flux}\label{sec:meth/calib_stellar_flux}

Our absolutely-calibrated stellar flux measurements closely followed the methodologies of \cite{Gordon2025}, though we used the \texttt{Eureka!}\ data analysis pipeline \citep{bell:2022} to do our reduction. We started from the MAST-provided \_uncal.fits files, removed any unflagged cosmic rays with a double-iteration 5$\sigma$ clipping of each pixel's time-series, and then replaced all bad pixels with bi-linear interpolation. We then used the \texttt{mgmc} centroiding method within \texttt{Eureka!}\ to measure the centroid of the median integration, using the expected source position from the FITS header as our starting guess of the centroid. We then used \texttt{Eureka!}'s \texttt{photutils} circular aperture+annulus extraction method with pixels being weighted by their fractional area that falls within the aperture/annulus (i.e., the `exact' \texttt{aperture\_edge} setting). Following \cite{Gordon2025}, we used a source aperture with radius of 5.69\,px and a background annulus spanning 8.63--11.45\,px. Within \texttt{Eureka!}'s Stage 4cal, we applied the aperture correction factor, $A_{\rm cor}$, of 1.497 as computed by \cite{Gordon2025} and used the only the in-eclipse measurements to compute the stellar flux, and compute the final measurement and uncertainty for each observation using the median and standard-deviation-of-the-mean.

Since we are treating each observation of TRAPPIST-1 as one multi-epoch time series, we do not need to inflate our uncertainties by adding in quadrature the $\sigma(\rm CF)$ error-inflation term from \cite{Gordon2025} as systematic uncertainties in the absolute calibration would be the exact same for all F1500W observations of TRAPPIST-1. However, it is unclear to what extent the $\sigma(\rm repeat) = 0.45\%$ systematic error term from \cite{Gordon2025} should be applied for our treatment of these data. On one hand, because our observations are all time-series observations and the mid-eclipse time for each visit is well after the start of the observations, they will be less affected by different levels of persistence from whatever MIRI filter was previously in-place and whatever field MIRI happened to be pointed towards (since MIRI doesn't use a shutter and is exposed to the sky at all times), and the $\sigma(\rm repeat)$ term may not apply to our situation. On the other hand, the $\sigma(\rm repeat)$ systematic error term from \cite{Gordon2025} may instead be dominated by variations in the PSF due to changes in the phasing of each of the different mirror segments between different visits, in which case $\sigma(\rm repeat)$ would equally apply to our analyses. Ultimately, the answer is unclear and further study into the cause of this systematic uncertainty is needed, so we optimistically chose to ignore the $\sigma(\rm repeat)$ term for our demonstrative paper.

\subsubsection{Combined visits reduction}\label{sec:meth/global_reduc}

The precision of the MIRI imaging mode for absolute flux measurements is limited (up to $\simeq$1\% of calibration uncertainty and approximately 0.1-1.2\% for repeatability \citep{Gordon2025}). This can make comparisons of absolute fluxes obtained months or years apart hard to compare. To mitigate these limitations we proceed to a joint reduction of various visits. We considered observations from the GTO program 1177 (PI: Greene) and GO~2304 (PI: Kreidberg), both obtained with MIRI imaging in the F1500W filter in FULL array mode over a 40-day period in fall 2022. We re-reduced the MAST pre-calibrated JWST data (calints.fits files) from both programs as if they belonged to a single observation that contained large gaps in between series of data points. This approach enables the computation of relative fluxes in between the 10~visits. We then converted the relative fluxes into absolute stellar fluxes by scaling them to the synthetic SPHINX stellar model, using as a reference the epoch when the observed flux is dominated by the star and the planetary contribution is negligible. This notably occurs when TRAPPIST-1 c is in secondary eclipse while TRAPPIST-1 b is also close to secondary eclipse ($BJD_{TDB}$ = 2459880.1872, \citep{zieba:2023}).

As the data were already pre-calibrated, we used \texttt{Eureka!} starting from Stage 3 to perform aperture photometry, and Stage 4 to generate the light curves. In Stage 3, we adopted an aperture radius of 6 pixels. For each integration, we recorded the center and width of the point spread function (PSF) in both x and y directions by fitting a 2D Gaussian. The background was estimated using an annulus spanning 15 to 25 pixels centered on the target and subsequently subtracted. The extracted fluxes were then processed in Stage 4, where we applied a 5$\sigma$ clipping to remove outliers, defined relative to a median flux computed with a 10-integration-wide boxcar filter.
Finally we compared these relative flux measurements to the phase curve model of the whole star+planets system. We set a reference flux calculated from the SPHINX spectrum to a time where the planets contribution is negligible (at mid-eclipse time of the first visit) such that only the star's flux is contributing.

\section{Results and Discussions}\label{sec:results}
\subsection{Modelled total phase curve}
\begin{figure}[h!]
    \centering
    \includegraphics[width=\linewidth]{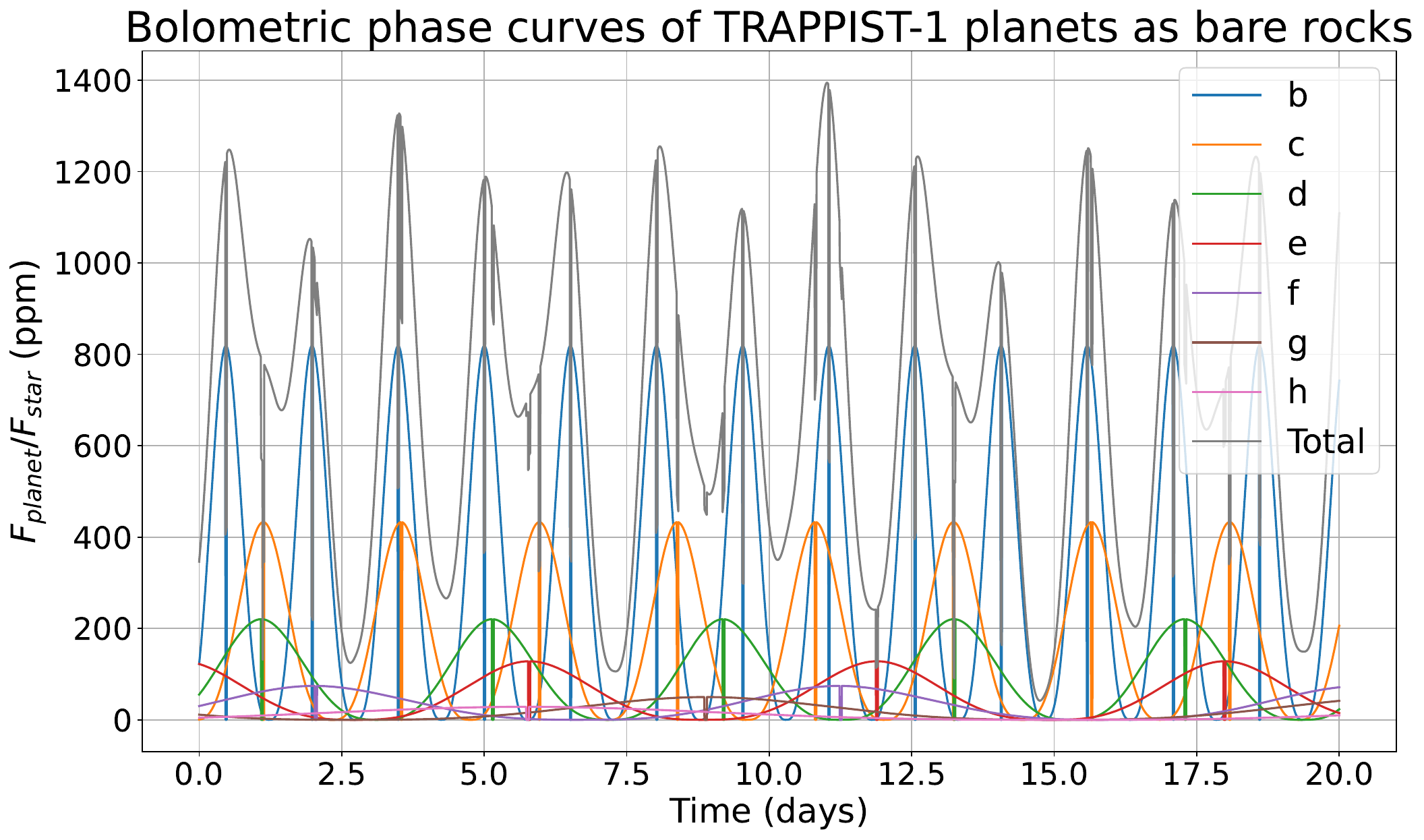}
    \caption{Bolometric flux emitted by each planet, and the resulting phase curve. Eclipses are displayed but not transits.}
    \label{fig:phc_total}
\end{figure}

The contributions of the different planets in the case of all bare rocks is displayed in Fig.~\ref{fig:phc_total}. As expected, planets b and c dominate the flux, but the contribution of the outer planets, especially d and e when in phase together, can add up to a flux comparable to TRAPPIST-1~c, typically detectable with JWST, though making this detection would require substantial observing time.

\subsection{Outer-planet contributions during archival JWST observations}\label{sec:outer_planets_contributions}

\begin{figure*}[p]
\centering
\rotatebox{90}{%
    \begin{minipage}{\textheight}
        \centering
        \includegraphics[width=\textwidth]{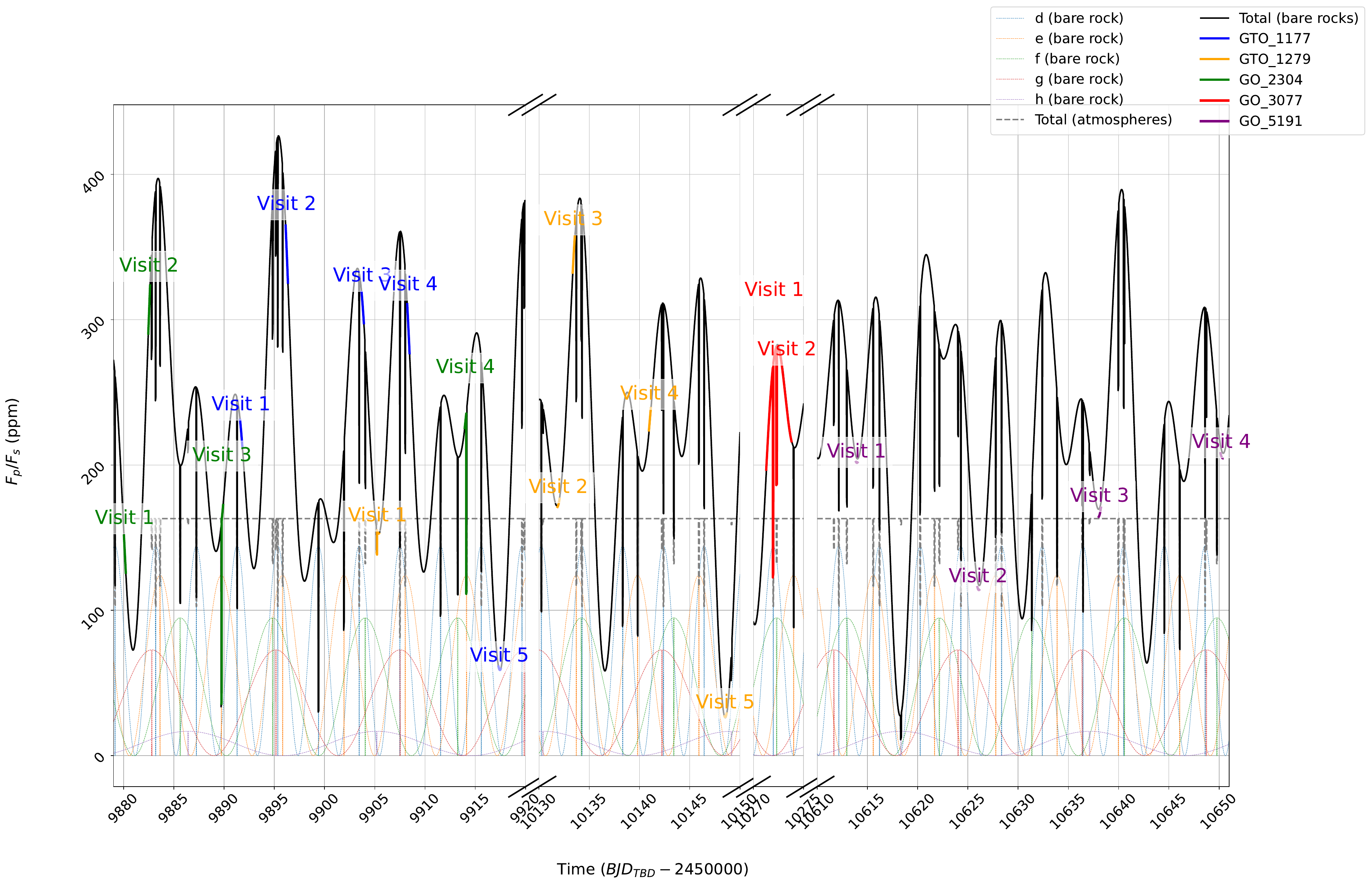}
        \caption{Simulations of the external planets' (d to h) contributions at the times of JWST observations (represented in color over the bare rocks total curve). Relative fluxes were simulated using the SPHINX spectrum at 12.8 microns. The contributions of each planet are shown in colored dotted lines and the case of all planets with thick atmospheres in dashed grey. The total phase curves correspond to the total contribution of the external planets. Eclipses are displayed but not transits. Visit~1 of GTO~1279 is not displayed as the observational strategy from this visit was different from the others and less precise \citep{ducrot:2024}.}
        \label{fig:simu_defgh_F1280W_sphinx_ppm}
    \end{minipage}
}
\end{figure*}

Fig.~\ref{fig:simu_defgh_F1280W_sphinx_ppm} shows the phase curve of the outer planets only (for better clarity), along with all JWST visits. The observations occur at different moments of contributions of the outer planets, allowing in theory to detect a difference between the atmospheric scenario and the bare rock one. Especially, visits 2 and 5 of GO~1177 happen in very different configurations, displaying the largest difference ($\sim$500~ppm) in the bare rock case (for the same wavelength filter).

\subsection{Observed fluxes and error bars}

Finally, we compared the modelled phase curve for the case where the flux variations are the largest (when all seven planets are bare rocks) to the measured flux at each visit, for the two different data analysis methods. 

\begin{figure}[ht!]
    \centering
    \includegraphics[width=\linewidth]{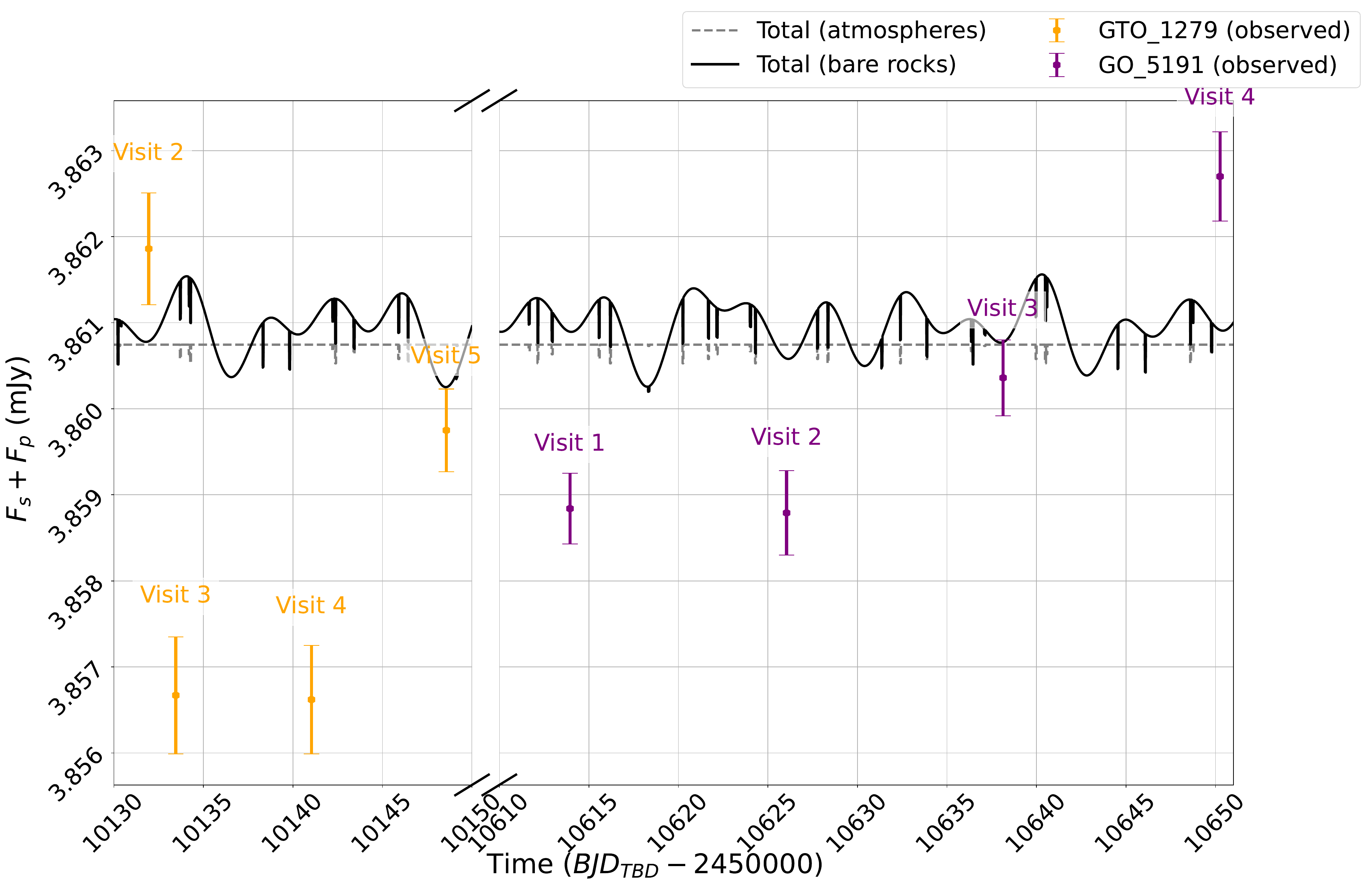}
    \includegraphics[width=\linewidth]{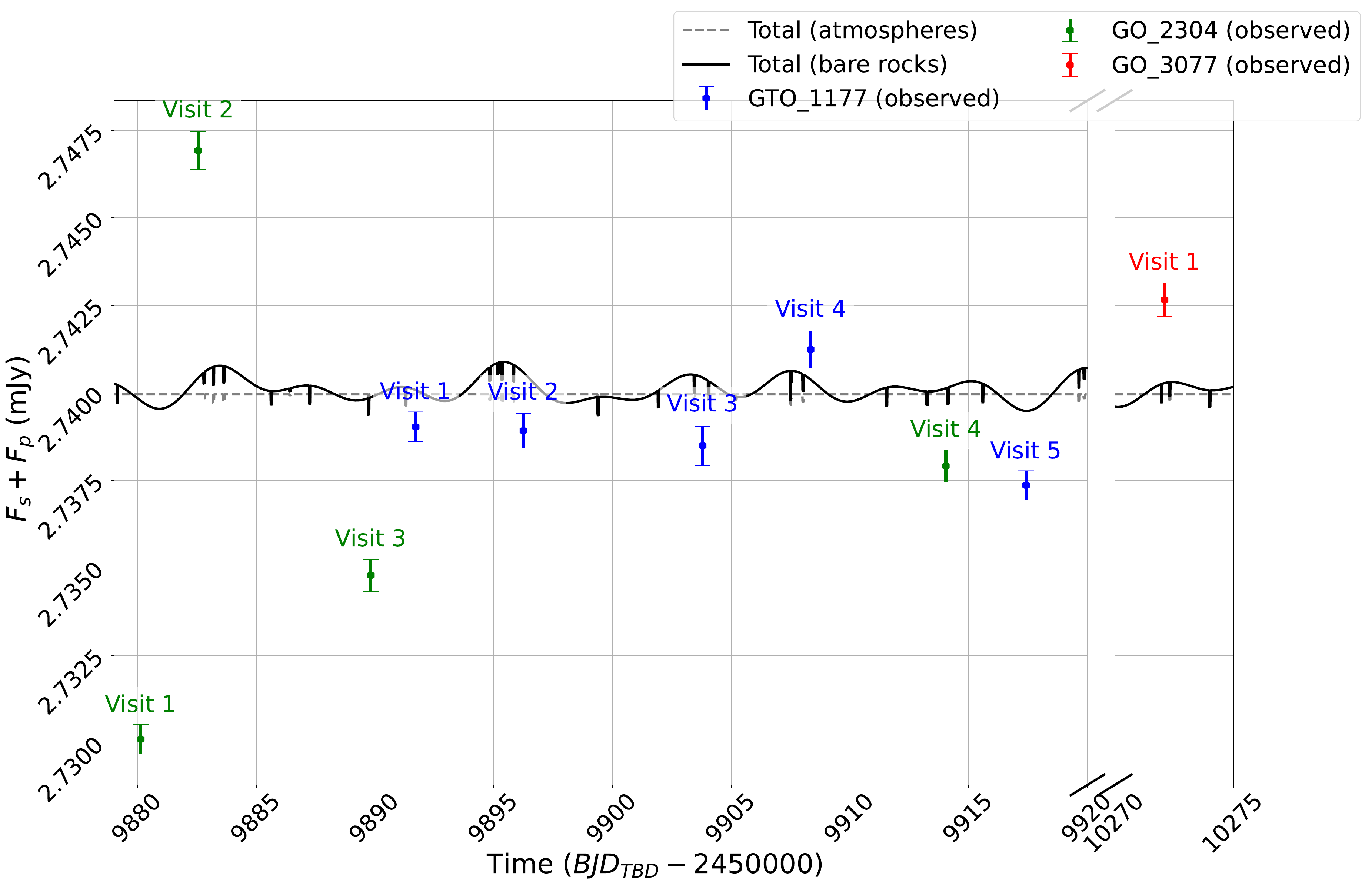}
    \caption{Modelled phase curve of the outer planets (d to h), along with absolute flux observations from every visit of JWST, using the SPHINX spectrum. Top panel: 12.8~\micron\ measurements; Bottom panel: 15~\micron. Visit~1 of GTO~1279 is not displayed as the observational strategy from this visit was different from the others and less precise \citep{ducrot:2024}. Offsets were applied to the observational points to place them at the same level as the curve: $+0.04$ mJy for GTO~1279, $+0.01$ mJy for GO~5191, $-0.13$ mJy for GTO~1177 and GO~2304, and $-0.06$ mJy for GO~3077.}
    \label{fig:obs_comparison}
\end{figure}

First, we measured the stellar flux in mJy for each visit using the method detailed in \S\ref{sec:meth/calib_stellar_flux} and compared it to the predicted fluxes; the results are shown in Fig.~\ref{fig:obs_comparison}. We see that the typical uncertainties on the total fluxes measured are significantly larger than the flux variations expected from the outer planets if they were bare rocks. The standard deviation of the measurements at 12.8 \micron\ is $16.9~\mu$Jy, and $21.2~\mu$Jy at 15 \micron. By comparison, the maximum flux variation for the planets thermal emission is expected to be $1.5~\mu$Jy and $1.6\mu$Jy at 12.8 \micron\ and 15 \micron\ respectively. This result highlights the fact that MIRI imaging is not precise enough in absolute flux estimation to detect the thermal flux variations of temperate rocky planets. 

\begin{figure}[ht!]
    \centering
    \includegraphics[width=\linewidth]{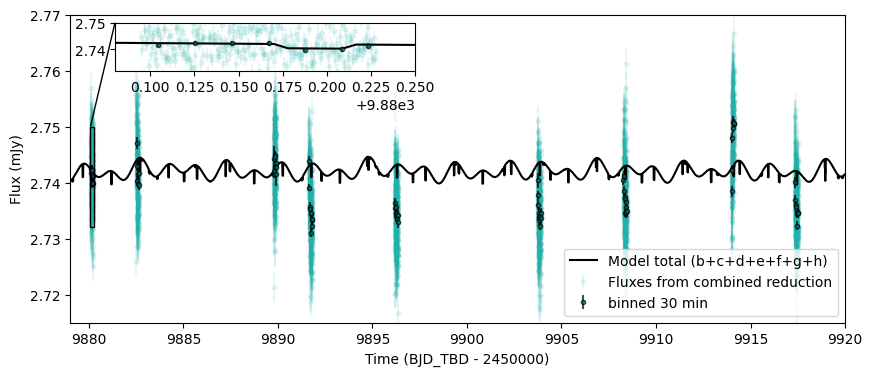}
    \caption{Fluxes versus time obtained from the combined reduction of all visits at 15 $\mu m$ from GTO~1177 and GO~2304 compared to the flux variations expected from the modelled total phase curve}
    \label{fig:combined_fluxes}
\end{figure}

Secondly, we compared the seven bare-rock planet model to the relative fluxes measured from broadband photometry on all visits treated as one single observation with gaps in the dataset; the results are shown in Figure~\ref{fig:combined_fluxes}. As detailed in \S\ref{sec:meth/global_reduc}, we could only apply this method at 15~$\mu m$ for GTO~1177 and GO~2304 as these were the only two programs that used the exact same strategy and same filter. We scaled the measured flux for the first visit of program GTO~1177 to be equal to the model prediction at the exact time of the mid-eclipse, see zoomed box on Figure~\ref{fig:combined_fluxes}. The fluxes from the other visits are then just computed using the relative flux scaled to this value. We observe that the precision is better when we processed with this relative flux approach. This time the dispersion in measured flux is $5.23\times10^{-5}$ mJy, one order of magnitude smaller than with absolutely-calibrated fluxes. Nevertheless, this precision in flux is still insufficient to detect the thermal flux variations from the outer planets for the 10~visits considered. 

\section{Perspectives}
\subsection{Limits}
The limits of this work lay in the difficulty of obtaining precise absolute flux measurements of the star+planets. First, since we are interested in the variation of the light of the total system (8 body problem that never goes back to the exact same configuration), we cannot stack data as we do when computing dayside planetary flux by stacking eclipse observations.
Second, the MIRI observations have an exquisite quality when it comes to computing relative flux and obtaining precise eclipse depths \citep{greene:2023, zieba:2023, ducrot:2024} or phase curve properties \citep{ducrotgillon:2024}, but the calibrated absolute stellar flux presents irreducible error bars. Our first reduction method consisting of working with the absolute fluxes is thus insufficiently precise.
On the other hand, the measurements must be taken with the same filters and the same sub-array as signals in observations taken with different subarrays can be lower by up to 3.4\% compared to FULL frame \citep{Gordon2025}. And even after correcting for the time and subarray dependencies, \cite{Gordon2025} computed that formal uncertainties on the flux calibration averaged for all observations are 0.3 to 1.0\% which is larger than the maximum absolute flux variations expected for the seven planets in the scenario where they are all bare-rocks ($\simeq$0.1\%).
The only solution would be to observe long continuous time-series with MIRI ($\simeq$50~hours) at specific phases when the sum of the outer planets contributions moves from one extrema (minimum or maximum) to the other. Indeed, the maximum contribution of the outer planet can reach 400-500~ppm in relative flux at times (see \S\ref{sec:outer_planets_contributions}) and the error on the relative flux binned in 30~minute intervals with MIRI F1280W and F1500W with the SUB256 subarray are 125 ppm and 136 ppm respectively. This approach and its feasibility is discussed in the following section.


\subsection{Long-duration continuous observations with MIRI imaging}
\label{sec:obs_predictions}

We thus used the \texttt{Exoplanets\_Phase\_Curves} code to predict the times when the variation of the flux would be at its maximum, by considering the flux difference between two consecutive extrema. Starting from 1st June 2026, we found 30~times when the variation of the flux of the external planets would be higher than 400~ppm. These times are listed in Table~\ref{table:flux_variations} and it is possible to notice that the typical lapse of time during which these large variations occur (i.e. between two consecutive extrema) is approximately 60~hours. 
However, the double phase curves of TRAPPIST-1~b and c from program GO~3077 showed that observations this long with JWST are possible \citep{gillon_double_2024}.

\begin{table}[h!]
\centering
\begin{tabular}{ccc}
$t_0$ (BJD\_TBD - 2450000) & Variation (ppm) & Duration (hours) \\
\hline
11194.9600 & 467 & 61.8006 \\
11219.5253 & 410 & 60.4806 \\
11231.4704 & 478 & 62.6406 \\
11253.5706 & 413 & 59.5206 \\
11267.9908 & 478 & 63.1206 \\
11290.0660 & 434 & 60.2406 \\
11304.5161 & 467 & 63.6006 \\
11326.5713 & 448 & 60.7206 \\
11341.0465 & 445 & 63.8406 \\
11363.0817 & 453 & 61.0806 \\
11375.0818 & 422 & 60.0006 \\
11377.5819 & 411 & 63.8406 \\
11399.5971 & 447 & 61.2006 \\
11411.5722 & 450 & 61.0806 \\
11436.1174 & 432 & 61.0806 \\
11448.0726 & 470 & 62.0406 \\
11472.6428 & 407 & 60.4806 \\
11484.5879 & 479 & 62.6406 \\
11506.6781 & 419 & 59.7606 \\
11521.1083 & 478 & 63.2406 \\
11543.1785 & 440 & 60.4806 \\
11557.6337 & 465 & 63.6006 \\
11579.6839 & 452 & 60.9606 \\
11594.1640 & 440 & 63.8406 \\
11616.1942 & 455 & 61.3206 \\
11628.1944 & 428 & 60.1206 \\
11630.6994 & 405 & 63.8406 \\
11652.7146 & 448 & 61.2006 \\
11664.6847 & 455 & 61.3206 \\
11689.2350 & 431 & 61.0806 \\
\hline
\end{tabular}
\caption{Predicted flux variations of more than 400~ppm from the external planets starting from 1st June 2026 for 500~days. The $t_0$ correspond to the time of the extremum at which starts the variation. 30~flux variations of more than 400~ppm were detected by the code.}
\label{table:flux_variations}
\end{table}

\subsection{Other works to detect the presence of an atmosphere on rocky planets using Mid-IR observations}

Other studies have explored the possibility of detecting the thermal emission of rocky planets through phase-dependent flux variations in the mid-infrared with JWST, in particular for the non-transiting habitable-zone rocky planet Proxima b.

First, \cite{Kreidberg2016} proposed measuring the thermal phase curve of Proxima b using MIRI’s LRS to maximise photon collection. However, given the long orbital period of Proxima b (11.8 days), such a programme would be very demanding in terms of JWST observing time. Furthermore, in light of the MIRI imaging performance presented in this work for detecting the thermal emission of rocky planets, we argue that the MIRI F1500W filter, centred on a CO$_2$ band, is better suited than MIRI LRS, as has been also demonstrated in several other publications \citep[e.g.,][]{mansfield:2019, koll:2019b, redfield:2024,espinoza:2025,hammond:2025}.

Then, \cite{Snellen2017} proposed a spectral high-pass filtering technique applied to the phase-curve spectra of Proxima b obtained with MIRI MRS, enabling the detection of CO$_2$ within a few days. This method relies on targeting a specific spectral feature (here the CO$_2$ band) in the combined planet+star spectrum and subsequently disentangling it from the stellar contribution. According to \cite{Snellen2017}, a stability of 100~ppm over 10~hours is required, with simulations performed over 4 × 24 hours.
More recently, \cite{Deming2024} investigated the potential of MIRI MRS for transit and eclipse spectroscopy and revisited the feasibility of the \cite{Snellen2017} method using updated instrument performance estimates from real data. They reported a stability of 138 ppm over 19 hours near 8 $\mu$m, which is an encouraging result for the viability of the \cite{Snellen2017} approach.
Such methods could also be explored for TRAPPIST-1 and other rocky planetary systems.

Finally, \cite{Stevenson2020} and \cite{Mandell2022} proposed the Planetary Infrared Excess (PIE) technique to probe the atmospheres of both transiting and non-transiting planets. This approach relies on acquiring simultaneous broad-wavelength spectra and isolating the planetary contribution (particularly prominent in the mid-infrared) from the stellar spectrum. Applied to Proxima b, they showed that this method is not feasible with JWST alone, but could become accessible with $\simeq 100$ observing hours using the proposed MIRECLE mission \citep{Mandell2022}.
Building on this idea, \cite{Mayorga2023} applied the PIE concept to an idealised TRAPPIST-1 system (neglecting both system distance and stellar activity) to assess whether the infrared excess from the seven planets could be detected with JWST MIRI/LRS. Even in the most favourable case of airless planets (maximising thermal emission), the PIE technique was unable to reveal their presence, largely due to degeneracies with the semi-major axes. In contrast, an idealised low-noise MIRECLE-like mission would enable detection, at least providing a lower limit on the number of planets.

Overall, these mid-infrared approaches to detecting atmospheres on rocky planets are promising. However, continuous time-series observations with the MIRI F1500W filter over partial phase curves may offer another alternative; as presented in the previous section.

\section{Conclusions}\label{sec:conclusions}
We re-analysed the data of five JWST programs observing TRAPPIST-1 in two MIRI filters: F1500W and F1280W. We aimed at using these data to try to detect the contribution of the outer planets of the system and make conclusions on the presence of a thick atmosphere on them.

We modelled the emission of the TRAPPIST-1 system in two end member cases: either all the planets of the system are airless with a zero albedo, or the outer planets (d-e) have a thick, completely redistributive atmosphere. This model was produced with \texttt{Exoplanets\_phase\_Curves} and is easily generalisable to other systems, open source, and available online. We compared this synthetic emission with the available observations, first by using absolute flux computed at mid-eclipse for each visit, then by performing a global reduction of all the visit as if they belong to one single observation. We found that in both cases the uncertainty of the measurements is larger than the variability associated to our modelled total phase curve. Thus, the current data is insufficient to give any information on the presence of an atmosphere on the outer planets of the TRAPPIST-1 system. We proposed 60-hours-long windows for observations that could allow to detect a potential variation caused by the phase curve of the outer planets.

Finally, we verified in this study that the variations of mid-infrared flux of the system caused by the outer planets is negligible in the context of the previous observations. Thus, the outer planets' emission does not create a source of contamination in the observation in emission of the two inner planets.

\section{acknowledgements}
The authors thank Karl Gordon for helpful discussions regarding the absolute flux calibration of MIRI photometry and the potential sources of systematic uncertainties in those measurements. The authors thank Martin Turbet for his original idea about the project and useful discussions and optimism about data reduction.

\newpage
\bibliographystyle{mn2e}
\bibliography{references}

\begin{appendix}

\section{Model validations}\label{sec:app/model_valid}
In this appendix we provide some comparisons between the simulated phase curves and observations in order to validate our code.

\begin{figure}[h!]
    \centering
    \includegraphics[width=\linewidth]{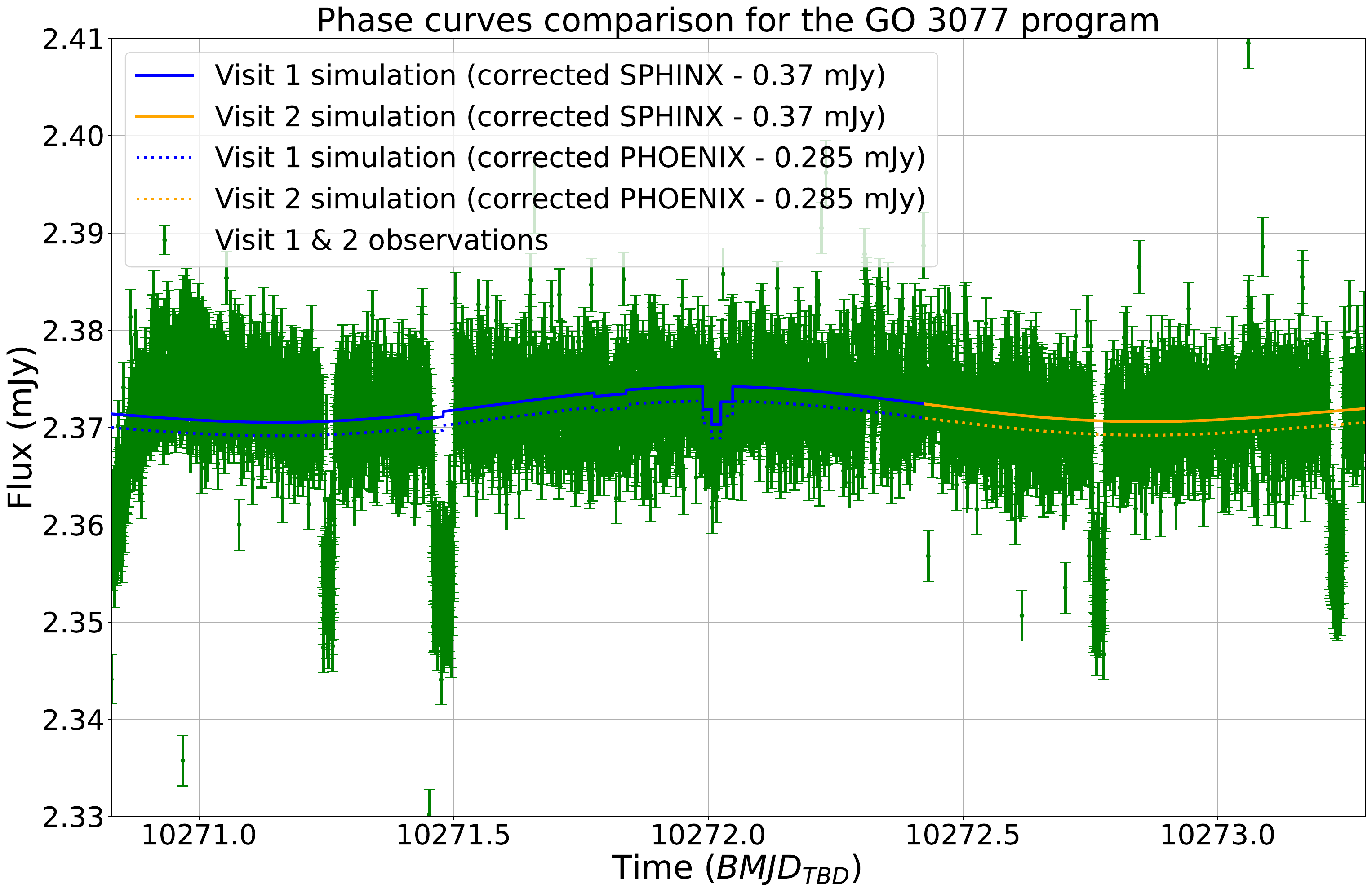}
    \caption{Comparison between the phase curve observed by the JWST during the GO~3077 program and the simulations of the total flux of star TRAPPIST-1 and its seven planets, using both the SPHINX and PHOENIX models. Offsets of $-0.37$ mJy and $-0.285$ mJy were respectively added to the simulations with the SPHINX and PHOENIX models in order to place these curves at the same level than the observations. The first and third transits correspond to planet b, the second one to g and the fourth to c. The transits are not displayed on the simulations. The ramp at the beginning of the observation comes from instrumental systematics.}
    \label{fig:GO3077_comparison}
\end{figure}

Figure~\ref{fig:GO3077_comparison} shows the observations of the GO~3077 program (total phase curve of the system) along with our simulation. We can see that the simulation matches the observations, except at the beginning where systematics dominate the observed signal.

Furthermore, comparison with the observations shows that the simulated phase curves place the eclipses at the correct timings. For example, in Figure~\ref{fig:simu_defgh_F1280W_sphinx_ppm}, eclipses of TRAPPIST-1 e are visible at the very beginning of visit 3 and at the very end of visit 4 of the GO~2304 program \citep{zieba:2023} which targeted eclipses of TRAPPIST-1~c but contained these eclipses of e. However, they are not visible in the data due to the too-low SNR, and observing eclipses of e would require a dedicated program.\\

\section{Phase curves with the PHOENIX model}

\begin{figure}[h!]
    \centering
    \includegraphics[width=\linewidth]{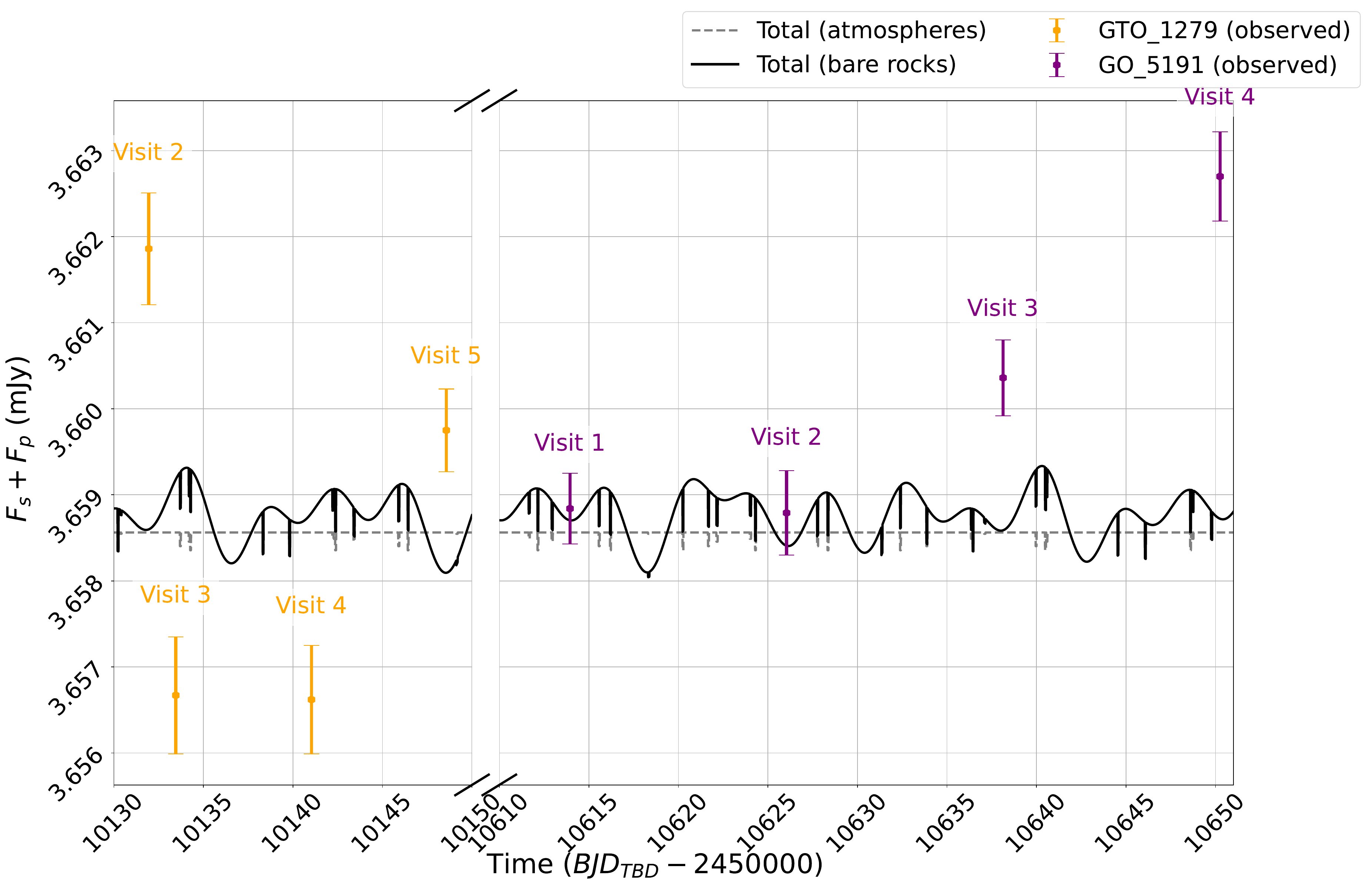}
    \includegraphics[width=\linewidth]{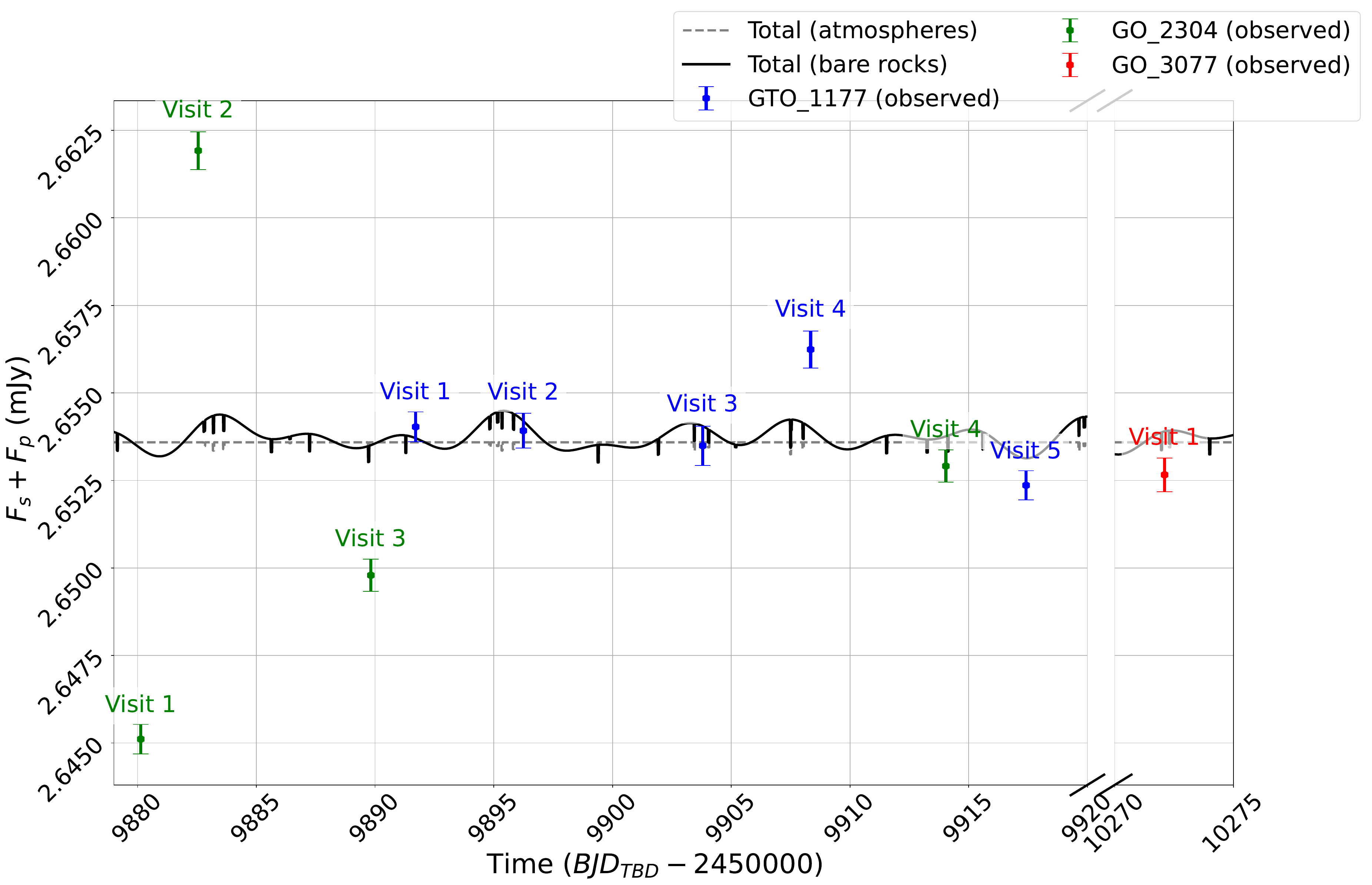}
    \caption{Modelled phase curve of the external planets (d to h), along with absolute flux observations from every visit of JWST, using the PHOENIX spectrum. Upper panel: 12.8~\micron\ measurements, lower panel: 15~\micron. Visit~1 of GTO~1279 is not displayed as the observational strategy from this visit was different from the others and less precise \citep{ducrot:2024}. Offsets were apply to the observation points to place them at the same level as the curve: $-0.16$ mJy for GTO~1279, $-0.19$ mJy for GO~5191, $-0.215$ mJy for GTO~1177 and GO~2304, and $-0.15$ mJy for GO~3077.}
    \label{fig:obs_comparison_2}
\end{figure}

The simulated total phase curve of the outer planets in Figure~\ref{fig:obs_comparison} used the SPHINX model for the stellar spectrum. We also used the PHOENIX spectrum \citep{allardModelsVerylowmassStars2012} for the phase curve simulation. The result is shown in Figure ~\ref{fig:obs_comparison_2}\\
The amplitude of the thermal flux variations using PHOENIX model is 1.42 $\mu$Jy at 12.8 \micron\ and 1.56 $\mu$Jy at 15 \micron, which is of the same order as with SPHINX and does not impact our conclusions.

\end{appendix}

\end{document}